\documentclass[letter]{aa}

\usepackage{graphicx}
\usepackage{hyperref}
\usepackage{natbib}

\newcommand{\cross}[1][1pt]{\ooalign{%
  \rule[0.7ex]{0.7ex}{0.15ex}\cr
  \hss\rule{0.15ex}{.5em}\hss\cr}}

\bibpunct{(}{)}{;}{a}{}{,}
\usepackage{txfonts}
\def\micron{$\mu$m}
\def\deg{$^\circ$}
\def\arcsec{"}

\def\aap{A\&A}

\def\apjl{ApJL}

\def\kms{${\rm km}\,{\rm s}^{-1}$}

\def\lesssim{\mathrel{\hbox{\rlap{\hbox{\lower4pt\hbox{$\sim$}}}\hbox{$<$}}}}
\def\gtrsim{\mathrel{\hbox{\rlap{\hbox{\lower4pt\hbox{$\sim$}}}\hbox{$>$}}}}

\def\kms{${\rm km}\,{\rm s}^{-1}$\,}

\def\msun{M$_{\odot}$\,}

\begin{document}
\title{The RCB star V854\,Cen is surrounded by a hot dusty shell}

\author{O. Chesneau\inst{1,\cross} \and F. Millour\inst{1} \and O. De
  Marco\inst{2} \and S. N. Bright\inst{1,2} \and A.~Spang\inst{1} \and
  E. Lagadec\inst{1} \and D. M\'ekarnia\inst{1} \and W. J. de
  Wit\inst{3} \thanks{Based on observations made with the VLTI at
    Paranal Observatory under program 091.D-0030 and
    093.D-0056.\protect \\ $^{\cross}$ O. Chesneau passed away shortly
    after submitting this letter. We express our profound sadness on
    this premature demise and convey our deepest condolences to his
    family.}  } \offprints{Florentin.Millour@oca.eu}

\institute{
Laboratoire Lagrange, UMR7293, Univ. Nice Sophia-Antipolis,
  CNRS, Observatoire de la C\^ote d'Azur, 06300 Nice, France              
	\and
Department of Physics \& Astronomy, Macquarie University, Sydney, NSW 2109, Australia
\and
European Southern Observatory, Casilla 19001, Santiago 19, Chile
}
\date{Received, accepted.}
\abstract
{}
{The hydrogen-deficient supergiants known as R Coronae Borealis (RCB)
  stars might be the result of a double-degenerate merger of two white
  dwarfs (WDs), or a final helium shell flash in a planetary nebula
  central star. In this context, any information on the geometry of
  their circumstellar environment and, in particular, the potential
  detection of elongated structures, is of great importance.  }
{We obtained near-IR observations of \object{V854\,Cen} with the {{\sc
      AMBER}} recombiner located at the Very Large Telescope
  Interferometer ({{\sc VLTI}}) array with the compact array
  (B$\leq$35m) in 2013 and the long array (B$\leq$140m) in 2014. At
  each time, \object{V854\,Cen} was at maximum light. The $H$- and
  $K$-band continua were investigated by means of spectrally dependant
  geometric models. These data were supplemented with mid-IR {{\sc
      VISIR}}/VLT images.  }
{A dusty slightly elongated over density is discovered both in the
  $H$- and $K$-band images. With the compact array, the central star
  is unresolved ($\Theta\leq2.5$\,mas), but a flattened dusty
  environment of $8 \times 11$ mas is discovered whose flux increases
  from about $\sim$20\% in the $H$ band to reach about $\sim$50\% at
  2.3\micron, which indicates hot (T$\sim$1500\,K) dust in the close
  vicinity of the star. The major axis is oriented at a position angle
  (P.A.) of 126$\pm$29\deg. Adding the long-array configuration
  dataset provides tighter constraints on the star diameter
  ($\Theta\leq1.0$\,mas), a slight increase of the overdensity to $12
  \times 15$ mas and a consistent P.A. of 133$\pm$49\deg. The closure
  phases, sensitive to asymmetries, are null and compatible with a
  centro-symmetric, unperturbed environment excluding point sources at
  the level of 3\% of the total flux in 2013 and 2014. The VISIR
  images exhibit a flattened aspect ratio at the 15-20\% level at
  larger distances ($\sim$1\arcsec) with a position angle of
  92$\pm$19\deg, marginally consistent with the interferometric
  observations. }
{This is the first time that a moderately elongated structure has been
  observed around an RCB star. These observations confirm the numerous
  suggestions for a bipolar structure proposed for this star in the
  literature, which were mainly based on polarimetric and
  spectroscopic observations. }

\titlerunning{The RCB V854\,Cen is surrounded by a hot dusty shell}
\authorrunning{Chesneau et al.}

\maketitle

\section{Introduction}
The R Coronae Borealis (RCB) stars are rare hydrogen-deficient
carbon-rich supergiants, best known for their spectacular declines in
brightness at irregular intervals \citep{2012JAVSO..40..539C,
  2002AJ....123.3387D}. Two evolutionary scenarios have been suggested
for producing an RCB star, a double-degenerate merger of two white
dwarfs (WD), or a final helium-shell flash in a planetary nebula
central star. However, the discovery that RCB stars have large amounts
of $^{18}$O is interpreted as a serious argument in favor of the
merger scenario \citep{2011ApJ...743...44C, 2007ApJ...662.1220C}. One
may speculate that the merger scenario leads to some observational
consequences, such as a fast rotation for the remnant star that leads
to a circumstellar environment with an axis of symmetry. Polarimetry
has been the main technique to reveal these symmetries, yet the
results have so far never been fully conclusive
\citep{2003A&A...412..405Y, 1997ApJ...476..870C, 1988ApJ...325L...9S}

\object{V854\,Cen} is an unusual member of the RCB class owing to its
relatively large hydrogen content and also to the PAHs detected in the
mid-IR (among other oddities). Extensive polarimetric observations
have been reported \citep{1993A&A...274..330K,
  1992AJ....103.1652W}. Another striking aspect of this star is the
fast wind that was spectroscopically detected, which reaches several
hundreds of \kms\ \citep{2013AJ....146...23C, 2003ApJ...595..412C,
  1999AJ....117.3007L, 1993MNRAS.264L..13C, 1993AJ....105.1915K,
  1992MNRAS.258P..33L}. The key aspect of the polarimetric
observations reported in \citet{1992AJ....103.1652W} is that the
emission lines were unpolarized, which implies that they were formed
outside regions of high dust
concentration. \citet{1993AJ....105.1915K} also provided strong
arguments that \object{V854\,Cen} may be at the origin of the bipolar
nebula based on high-resolution spectroscopy. HST long-slit,
far-ultraviolet spectra of \object{V854\,Cen} showed that the C\,II
emission region around \object{V854\,Cen} is significantly extended by
about 2.5\arcsec\ \citep{2001ApJ...560..986C}.

With the advent of optical interferometry, many dusty environments
around evolved stars were resolved
\citep{2011apn5.confE.218C,2009A&A...493L..17C}. Some first attempts
to monitor the dust production of the RCB \object{RY\,Sgr} at high
spatial resolution were performed with the MIDI/VLTI instrument
\citep{2007A&A...466L...1L}. A few observations of \object{V854\,Cen}
are reported in \citet{2011MNRAS.414.1195B}. But the limited $(u,v)$
coverage still hampered detecting clear departure from spherical
symmetry.

This letter presents optical interferometry measurements obtained with
the Very Large Telescope Interferometer (VLTI) and reports the discovery
of a moderately elongated structure around \object{V854\,Cen}. The
observations are presented in Sect. \ref{Observations}. In
Sect. \ref{Analysis} we analyze the {{\sc AMBER}} $H$, $K$ continuum
measurements by means of simple geometrical spectrally dependent
models together with {{\sc VISIR}} mid-IR images. The results are then
discussed in Sect. \ref{Discussion}.


\section{Observations}
\label{Observations}

\begin{table}[bhtp]
  \caption[]{Log of  \object{V854\,Cen}  {{\sc AMBER}} observations.}
  \label{tab:logObs}
  \centering
  \begin{tabular}{lcccccccc}
    \hline
    \noalign{\smallskip}
    Date & Stations &  Wavelength & Nb. Obs. \\ 
    \noalign{\smallskip}
    \hline
    \noalign{\smallskip}
    14/04/2013 & A1-B2-D0 &  1.54 -- 2.40 $\mu$m &  5 \\
    15/04/2013 & A1-C1-D0 &  1.54 -- 2.40 $\mu$m &  3 \\
    06/05/2014 & A1-G1-K0 &  1.54 -- 2.40 $\mu$m &  2 \\
    06/05/2014 & A1-G1-J3 &  1.54 -- 2.40 $\mu$m &  1 \\
    \noalign{\smallskip}
    \hline
  \end{tabular}
  \tablefoot{\tiny Calibrator angular diameters from SearchCal@JMMC
    \citep{2006A&A...456..789B}: HD\,124433 $0.916\pm0.065$\,mas,
    HD\,127214 $0.57\pm0.04$\,mas. }
\end{table}

\begin{figure}[htbp]
 \centering
 \includegraphics[width=6.cm]{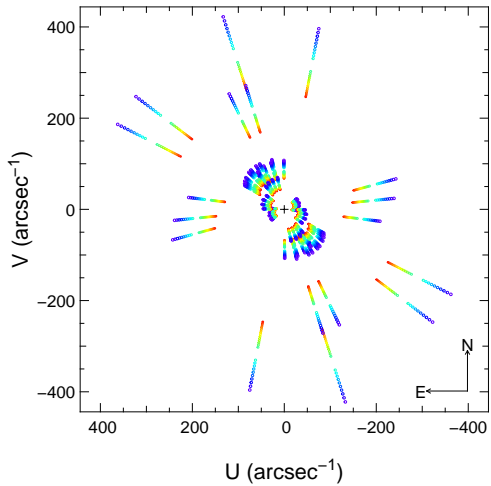}
 \caption{Spectrally dispersed $(u,v)$ coverage of the compact 2013 and
   extended 2014 configuration of the {{\sc AMBER}}
   observations. \label{fig:AMBER_uv}}
\end{figure}


The interferometric observations were obtained with the low spectral
resolution mode of {{\sc AMBER}} (R=35), a three-telescope combiner
located at the VLTI \citep{2007A&A...464....1P}. The observations were
performed with the 1.8m auxiliary telescopes (ATs) under photometric
conditions, using the compact configuration ($B\leq35$m) in 2013 and
the long configuration ($B\leq140$m) in 2014.  The observations log is
presented in Table\,\ref{tab:logObs} and the ($u,v$) plan coverage of
the compact configuration is plotted in Fig.\ref{fig:AMBER_uv}. We
reduced the data using the standard {{\sc Amber}} data reduction
software {\tt amdlib v3.0.3b1}
\citep{2009A&A...502..705C,2007A&A...464...29T}. The visibilities
secured in 2013 and 2014 are represented in
Fig.~\ref{fig:AMBER_Vis_UD}. Noteworthy, the American Association of
Variable Star Observers ({\sc AAVSO}) database shows that
\object{V854\,Cen} was at maximum light, V$\sim$7.5-7.2 when the 2013
and 2014 {{\sc AMBER}} observations were obtained.

The AMBER visibilities, a proxy of size and shape of the system, show
strong variations as a function of spatial frequencies and as a
function of wavelengths. This means that the object is resolved by
AMBER and that its shape spectrally varies between the $H$ and $K$
bands. The closure phases, a proxy of asymmetries in the image of the
system, are equal to zero within the uncertainties. This means that
this probably is a centro-symmetric object.

\begin{figure}[htbp]
 \centering
 \includegraphics[width=7.cm]{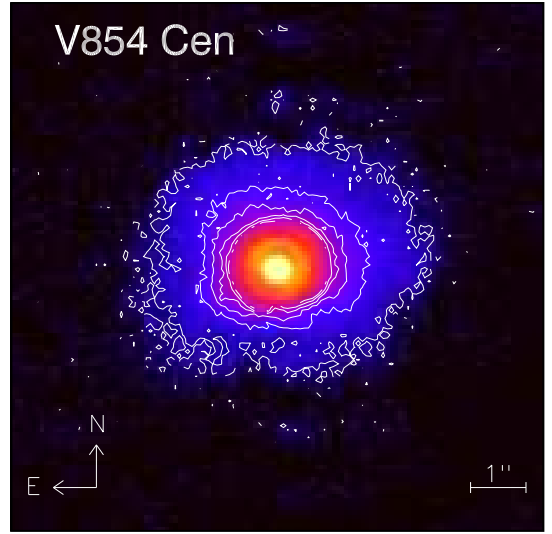}
 \caption{11.65$\mu$m VISIR image with enhanced contours of the
   external regions. \label{fig:VISIR_im}}
\end{figure}

\object{V854\,Cen} was observed on 30 of June 2008 with the
mid-infrared imager {{\sc VISIR}}/VLT \citep{2004Msngr.117...12L} as
part of an imaging survey of post-AGB stars and related objects
\citep{2011MNRAS.417...32L}. The images were obtained with a pixel
scale of 0.075\arcsec and a field of view of $19.2 \times 19.2$ arcsec
through the SiC filter ($\lambda=11.65\mu$m, $\Delta \lambda$
2.34$\mu$m). Narrow-band filter images were also obtained, but had a
too low S/N for scientific use. The data were processed as described
by \citet{2011MNRAS.417...32L}, with an additional step to remove
horizontal stripes produced by the mid-infrared detector, wor which we
used filtering in the Fourier space. Fig.~\ref{fig:VISIR_im} displays
an image of \object{V854\,Cen} obtained through the SiC filter at
11.65$\mu$m ($\Delta \lambda$ 2.34$\mu$m) for which the contrast of
the faintest regions was enhanced. The bright point-source is clearly
surrounded by a fainter structure with a largest extent of
$\sim$3\arcsec.


\section{Analysis}
\label{Analysis}

We analyzed the {{\sc AMBER}} data with the {\tt fitOmatic} routine
\citep{2009A&A...507..317M}, which enabled us to introduce different
spectra for different simple geometric components. Various
combinations of geometrical models were tested, including uniform
disks, Gaussian disks, rings, power-law and exponential-law-profile
disks. Our best-match model for the compact array is a two-component
model, consisting of an unresolved uniform disk ($\Theta\leq2.5$\,mas,
star component), and a flattened Gaussian (shell component) with a
FWHM of the minor axis of $8\pm 1$ mas, and a major axis of
$11\pm3$\,mas. The orientation of the major axis is
126$\pm$29\deg. But it must be kept in mind that the $(u,v)$ coverage
is not ideal for an accurate determination of the P.A. angle and
flattening of the structure. The quality of the fit is relatively good
with a reduced $\chi^2$ of 1.5. Combining the 2013 compact-array data
with the 2014 long-array data brings consistent new information. The
star remains unresolved ($\Theta\leq1.0$\,mas), and a slight increase
of the apparent diameter with a minor axis of $12\pm 2$ mas, and a
major axis of $15\pm5$\,mas is measured (reduced $\chi^2$ of 4.9). The
orientation of the major axis is 133$\pm$49\deg. The aspect ratio of
the shell is moderate, and we note that a round structure of
$\Theta=12.3\pm 2$ could also fit the data, albeit with a larger
reduced $\chi^2$ of 6.0. Assuming a disk-like geometry, the observed
flattening would correspond to an inclination of $i\leq60$\deg. The
spectra of both components are spatially and spectrally isolated. The
star flux relative to the total flux steadily decreases from 85\%
at $1.53\mu$m to 42\% at $2.49\mu$m, while the shell relative flux
increases from 15\% to 58\%, respectively, with photometric errors at
the level of 25\%. Assuming a temperature of 6750\,K for the central
star \citep{2011MNRAS.414.1195B}, a hot temperature $T\leq1400$K is
inferred for the shell, which explains its direct detection in the
near-IR.

\begin{figure}[htbp]
 \centering
\includegraphics[height=0.48\textwidth, angle=-90]{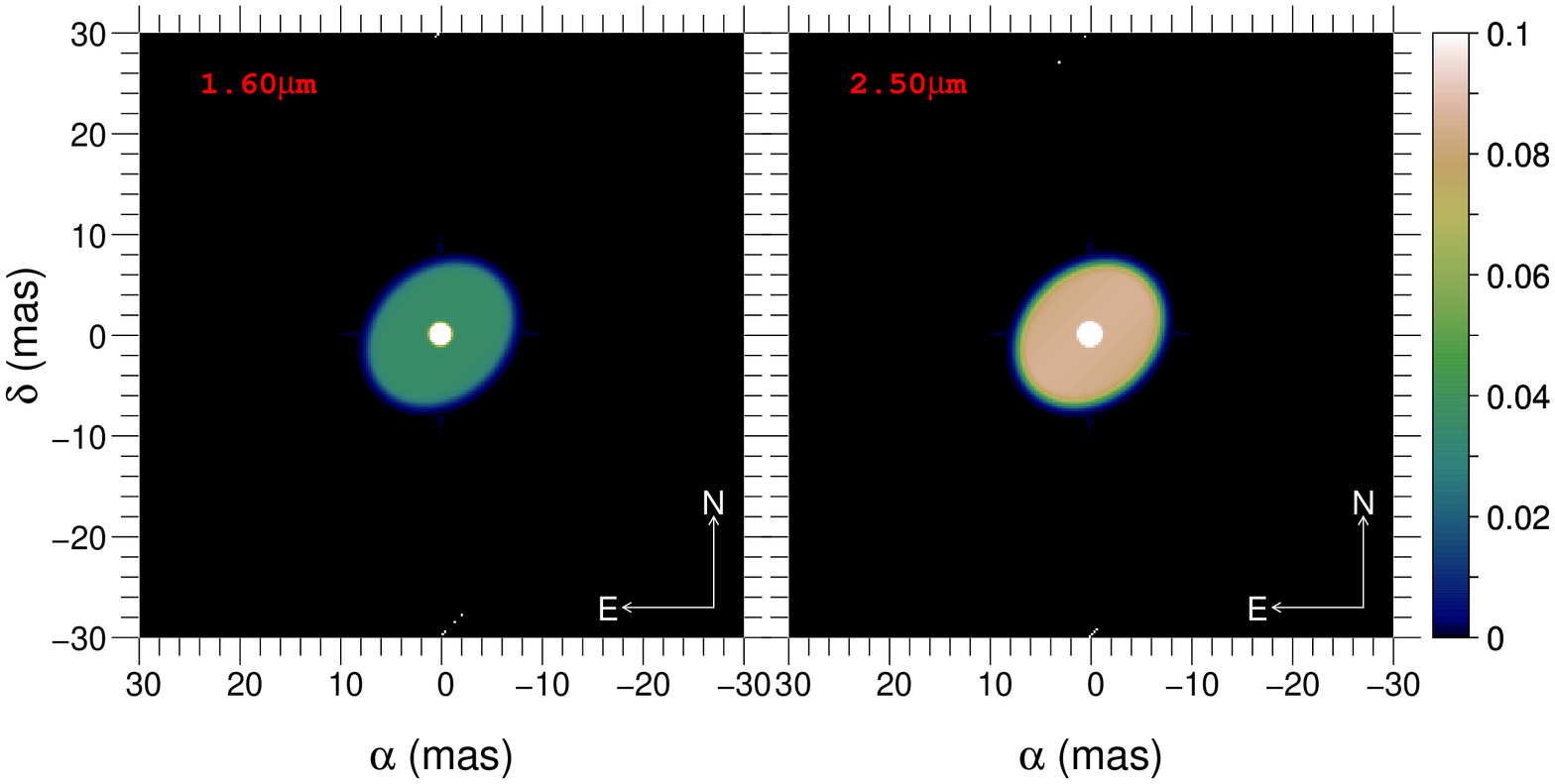}\\
\vspace{5mm}
\includegraphics[height=0.48\textwidth, angle=-90]{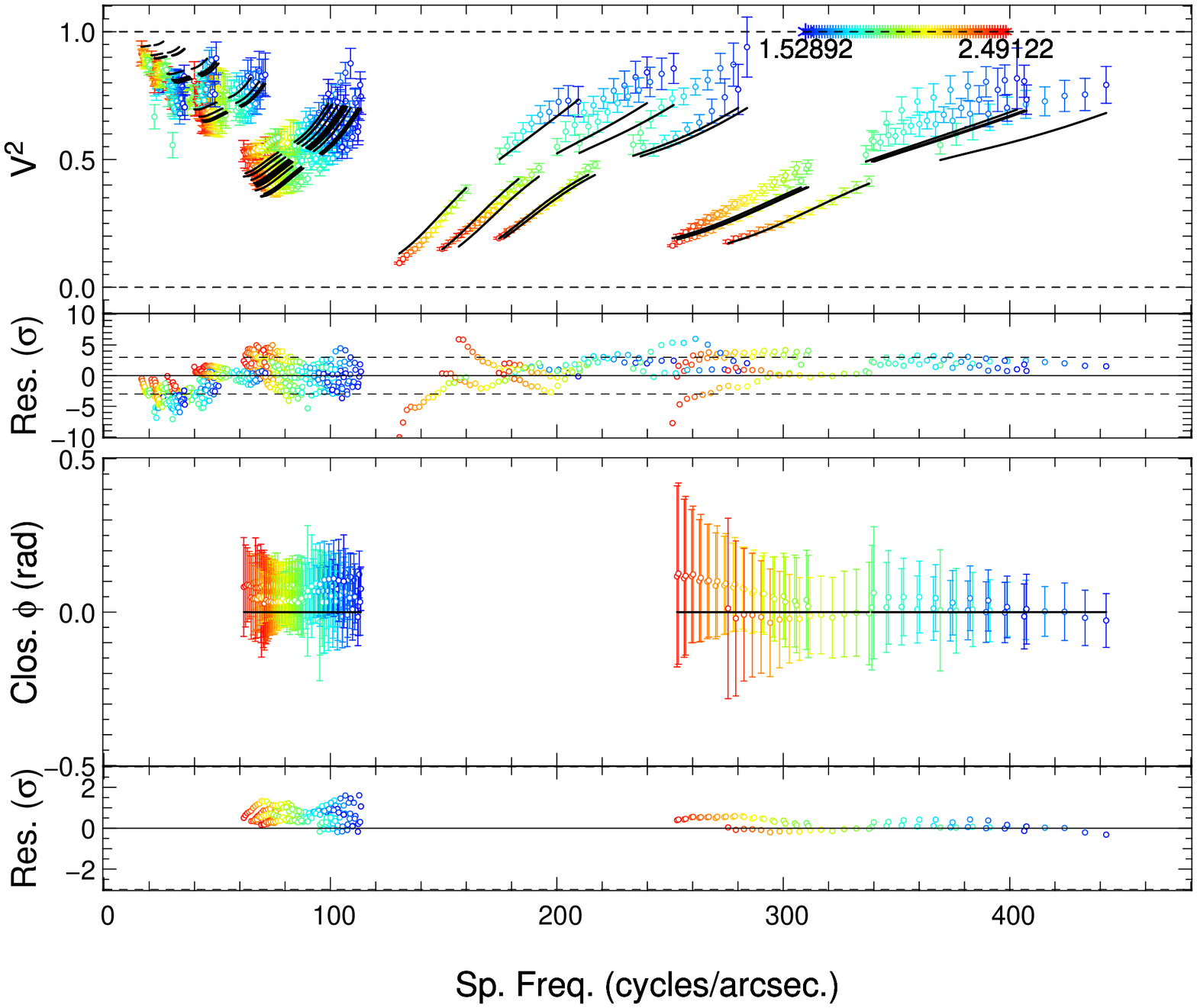}
\caption{{\bf Top:} Best geometrical models of the environment of
  V854\,Cen at two selected wavelengths. {\bf Middle:} 2013/2014
  combined $H$ (blue) and $K$ (red) bands interferometric visibilities
  obtained with {{\sc AMBER}}/VLTI compared with the signal from the
  polychromatic geometrical modeling. The wavelength label is in
  $\mu$m. {\bf Bottom:} Closure phases. Colors are the
  same. \label{fig:AMBER_Vis_UD}}
\end{figure}

\begin{figure}[htbp]
 \centering
\includegraphics[height=0.48\textwidth, angle=-90]{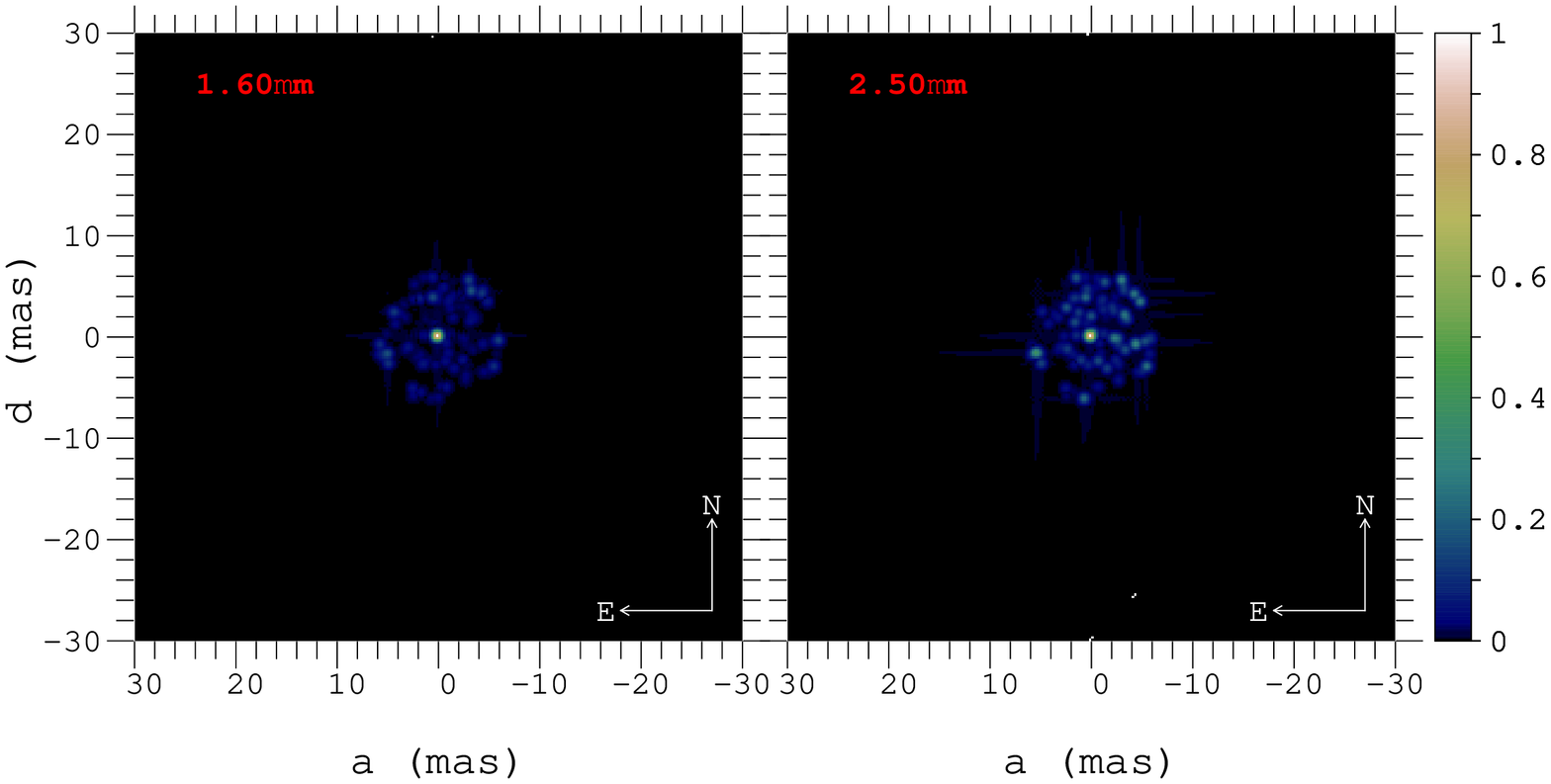}\\
\vspace{5mm}
\includegraphics[height=0.48\textwidth, angle=-90]{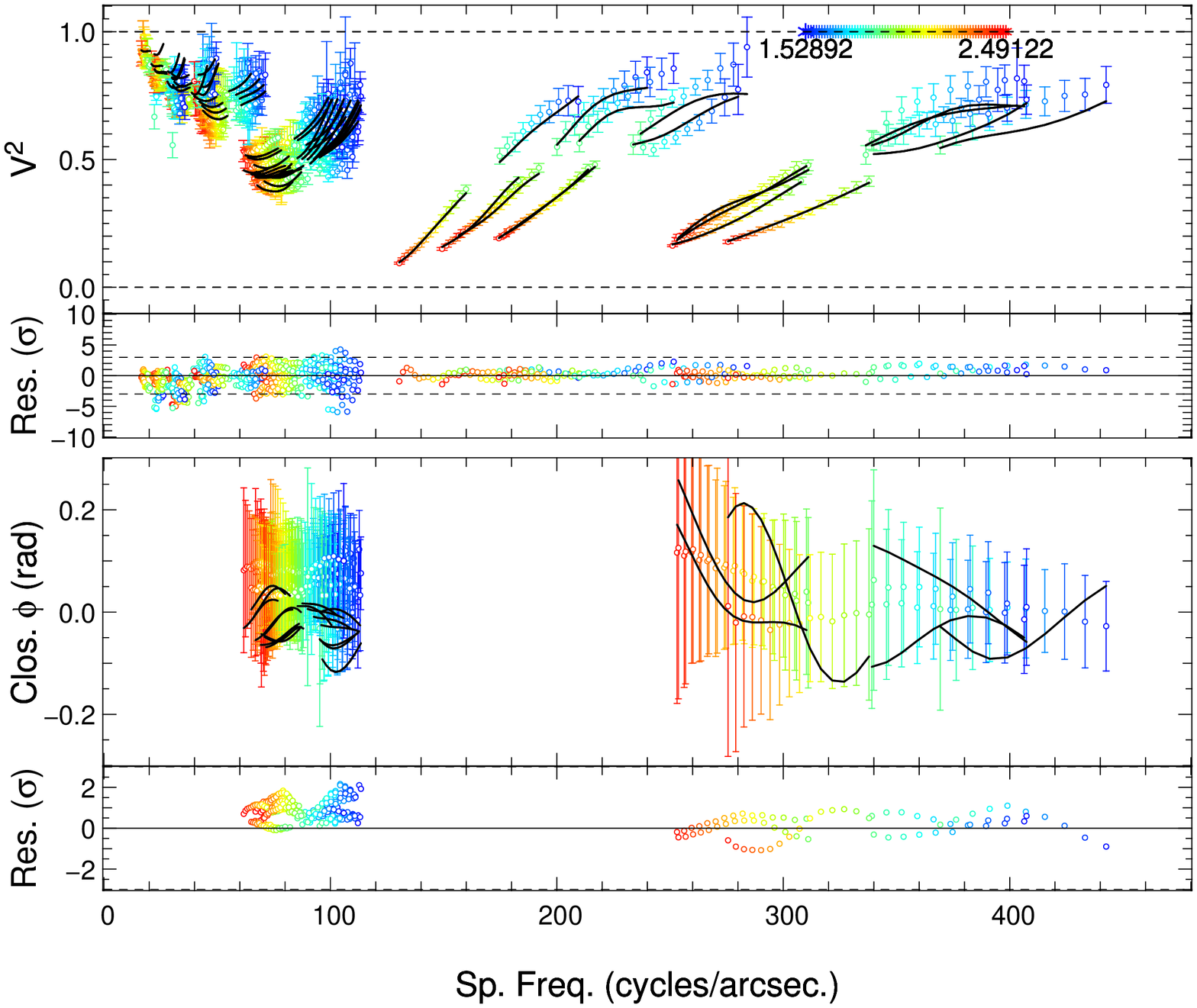}
\caption{{\bf Top:} One of the best models for a clumpy
  environment of V854\,Cen at two selected wavelengths with 30
  clumps. {\bf Middle:} 2013/2014 combined $H$ (blue) and $K$ (red) band
  interferometric visibilities obtained with {{\sc AMBER}}/VLTI
  compared with the signal from the polychromatic geometrical
  modeling. The wavelength label is in $\mu$m. {\bf Bottom:} Closure
  phases. Colors are the same. \label{fig:AMBER_Vis_clumps}}
\end{figure}

In Fig.~\ref{fig:AMBER_Vis_UD}, we show an illustration of the data
fit with our best geometrical model. We tested for a companion star or
a single clump of dust in the shell given the closure phase signal we
observed on \object{V854\,Cen} (null within the error bars). The test
showed point-like sources whose flux were always lower than 3\% of the
total flux, which means that this is well below the detection limit
for such asymmetries in the system with the current data down to a
typical spatial resolution of 10\,mas. We also tested a fully clumped
uniform shell that contained up to 30 clumps, to see whether the high
$H$-band visibilities could be explained this way, and this may indeed
be an explanation of these high visibilities, although we cannot
clearly infer the clump structure (because there are too many
parameters to fit). The clumpy model is shown in
Fig.~\ref{fig:AMBER_Vis_clumps}. Importantly, the $K$-band
visibilities (Fig.~\ref{fig:AMBER_Vis_UD}) at the highest spatial
frequencies show a striking increase, after a deep decrease of
visibilities near the spatial frequency of 100\,arcsec$^{-1}$. We
interpret this feature as a first visibility lobe followed by a second
visibility lobe that is clearly seen in the data. This is a typical
feature of objects whose brightness distribution contains a sharp edge
\citep[see, for instance, an example in][]{2009A&A...506L..49M}.
Indeed, such a sharp edge would produce wiggles in the Fourier
transformation of the image of the object (and hence in the
visibilities). This supports the idea that the shell around
\object{V854\,Cen} is slightly better described by a (truncated)
uniform disk (reduced $\chi^2$ of 4.9) than by a (smooth) Gaussian
disk (reduced $\chi^2$ of 7.0), hence the sharp edges in the
geometrical models shown in Fig.\ref{fig:AMBER_Vis_UD}).

The lower contours of the {{\sc VISIR}} image were analyzed to study
their geometry by fitting 2D Gaussian at different flux levels. This
yielded a good estimate of the nebula's ellipticity and orientation as
a function of the distance to its center. An elongated structure is
discovered with a P.A. of 92$\pm$19\deg\ and an ellipticity of
$\sim$1.2. The resolved parts of the image represent 5.4\% of the
total flux in the SiC filter.

This P.A. is only marginally constistent with the P.A. derived with
{\sc AMBER}. It is possible that the spatial structure of the mass
loss of V853 Cen is inhomogenous and randomly variable, which would
explain this difference.

We attempted to check the consistency of the P.A. determined by {{\sc
    AMBER}} and {{\sc VISIR}} and those detected by polarimetry. This
information is not systematically published, therefore we refer to a
few other publications. According to \citet{1993A&A...274..330K}, a
position angle of $\sim$65\deg\ is observed in the V band for maxima
episodes during which the direct starlight is less affected by dust
clumps and the polarimetric signal is lowest (such as at JD=2448299.0
with V=7.2). The U band is also highly polarized ($2.82\pm0.1\%$) with
a measured P.A. angle of $44\pm1$\deg. \citet{1993AJ....105.1915K}
also proposed a polarized P.A. angle of 65$\pm$15\deg\ that
corresponds to the scattering from a disk. Unfortunately,
\object{V854\,Cen} is very active and produces dust clumps at a high
rate, which affects the photometric bands differently and hampers
detecting a polarimetric signal from a stable circumstellar
environment.

\section{Discussion}
\label{Discussion}

We used a compact configuration (baselines limited to $\sim$50m)
optimized for studying the circumstellar environment of
\object{V854\,Cen}. A moderately flattened dusty environment was
discovered around \object{V854\,Cen}. This shed new light on the
evolutionary scenarios that have been suggested for the production of
RCB stars: the double-degenerate merger of two white dwarfs, which may
lead to an axis of symmetry that promotes equatorially enhanced
mass-loss and disks, or the final helium-shell flash in a planetary
nebula central star, which may retain a central symmetry and promote
no disk \citep{2011ApJ...743...44C, 2007ApJ...662.1220C,1984ApJ...277..355W}.

Below we elaborate further on the importance of this discovery within
the context of RCB star formation scenarios. What is the origin of
this slightly elongated environment? The closure phases are close to
zero, which excludes a close companion or single dust clump in the
vicinity of the star during this period of maximum light that was
brighter than 3\% of the total flux, that is, a flux difference of at
least 4 magnitudes. This may imply that the dust is distributed in a
disk that possibly is the relic of a past event, or that the observed
spatial distribution reflects a field of randomly launched clumps in
free-fall around the central star. The disk, or circumstellar
structure, which would be constantly replenished by a discrete or
continuous process, would explain the very hot temperature of the
dust. An inspiring comparison may be made with the so-called Be stars
\citep{2013A&ARv..21...69R}, for which the fast rotation of the
central star together with its pulsational properties triggers the
formation of a dense circumstellar disk. But, this does not imply that
the dust clumps are exclusively formed in the equatorial plane,
because if they were, the light declines would be much less frequent
this very active star. Furthermore, even though the original merged
star might have been rotating quickly, when it became a giant and
became larger by a factor of about $10^4$, the surface speed would
have dropped by that factor. The RCB stars are known to be slow
rotators, and the narrow emission lines of \object{V854\,Cen} are
unresolved \citep[$\leq 20$\kms,][]{1993AJ....105.1915K}.
\object{V854\,Cen} has a well-known single-pulsation period (43.2
days) whose phase is related to the formation of the dust
\citep{2007MNRAS.375..301C}. Convection may also contribute to the
launching process of the dust clumps, like in cool supergiant stars.

The intermediate inclination of the system (we recall that $i \leq
60$\deg) implies that the dust clumps that are regularly launched must
originate from the high latitudes of the star to intercept the line of
sight and dim its visual flux this deeply! This indicates a non radial
formation and launching process of the carbonaceous dust clumps.

The polarization properties of \object{V854\,Cen} are similar (albeit
weaker) to those of the intermediate-luminosity red transient (ILOT)
\object{V4332 Sgr} \citep{2013A&A...558A..82K,2009ApJ...699.1850B}
that suffered an outburst in 1994 and later developed the
characteristics of a large giant. ILOTs are outbursts with energies
intermediate between those of novae and supernovae
\citep{2012PASA...29..482K} examples of which are V838 Mon
\citep{2003Natur.422..405B} or V1309 Sco
\citep{2011A&A...528A.114T}. \citet{2013A&A...558A..82K} show that the
polarized continuum of \object{V4332 Sgr} in 2011 disagrees with the
unpolarised emission lines. This is interpreted as evidence for a disk
seen at high inclination that completely obscures the central
source. The star light would then reach us after scattering on dust
located above and below the disk. The emission lines, on the other
hand, would derive from photoexcitation by stellar radiation of the
molecules located in the material above and below the plane. This
configuration is also observed for the \object{Sakurai object}
\citep{2014ApJ...785..146H,2009A&A...493L..17C}, another object
thought to have suffered a very late final helium flash. Noteworthy,
strong abundance similarities between \object{Sakurai object} and
\object{V854~Cen} sources were detected \citep{1998A&A...332..651A}.

If the merger scenario for the RCBs applies, then they too may be in
the ILOT range if observed at the time of the merger. Two
0.5\msun\ WDs will deliver quite a substantial amount of gravitational
energy ($\lesssim 5^{49}$ erg), more than would be the case for a
main-sequence star merger, therefore they may cluster in the upper
region of the ILOT locus on the energy-time diagram. The stellar
expansion that ensues would make the object a giant, as is the case
for ILOTs. WD-WD merger simulations of RCB stars are still relatively
crude \citep{2012ApJ...757...76S} and do not simulate the possible
formation of a disk. However, the post-merger is a fast rotator before
the radial expansion. Magnetic fields, which were not included in the
simulations probably play an important role, because they potentially
affect many aspects of the star activity, such as its pulsations and
whether an axial-symmetry is established. An interesting possibility
is that the dust clumps launched by the star from its high latitudes
are not able to ballistically leave the system and accumulate in the
equatorial plane in a process reminiscent of the so-called
wind-compressed disk proposed by \citet{1993ApJ...409..429B}. The
launching conditions of dust around cool stars have recently been
investigated in depth by \citet{2009ASPC..414....3H,
  2007ASPC..378..145H}.  More observations are needed to link the
extended nebula detected by {{\sc VISIR}} and the compact structure
resolved with the VLTI. High angular resolution ($\sim$20-50mas)
polarimetric and coranagraphic imaging in the optical such as provided
by the {{\sc SPHERE}}/VLT instrument will be a key to achieve this
goal and better understand the dust ejection mechanism in
\object{V854\,Cen}.

\begin{acknowledgements}
  A great thanks for the ESO staff for the help with these
  observations, and in particular Thomas Rivinius. We also thank the
  referee, Pierre Kervella, for his help in improving this paper. This
  work has made use of JMMC, CDS and AAVSO resources.
\end{acknowledgements}

\bibliographystyle{aa}
\bibliography{RCrB}

\end{document}